\documentclass[letterpaper, 10 pt, conference]{ieeeconf}  

\IEEEoverridecommandlockouts                              
\usepackage{amsmath}
\usepackage{amsfonts}
\usepackage{amssymb}
\usepackage{bigints}
\usepackage{graphicx}
\usepackage{epstopdf}
\usepackage{subcaption}
\usepackage{booktabs}
\usepackage{framed}

\newcommand{\rar}{\rightarrow}

\newcommand{\be}{\begin{equation}}
\newcommand{\ee}{\end{equation}}
\newcommand{\bea}{\begin{eqnarray}}
\newcommand{\eea}{\end{eqnarray}}
\newcommand{\bes}{\begin{eqnarray*}}
\newcommand{\ees}{\end{eqnarray*}}

\newcommand{\bi}{\begin{itemize}}
\newcommand{\ei}{\end{itemize}}
\newcommand{\ben}{\begin{enumerate}}
\newcommand{\een}{\end{enumerate}}

\newcommand{\bp}{\begin{problem}}
\newcommand{\ep}{\end{problem}}
\newcommand{\hso}{\hspace{.1in}}
\newcommand{\hst}{\hspace{.2in}}

\newcommand{\noi}{\noindent}

\newcommand{\bc}{\begin{center}}
\newcommand{\ec}{\end{center}}

\title{\LARGE \bf
Numerical Evaluation of Exact Person-by-Person  Optimal Nonlinear Control Strategies of the Witsenhausen Counterexample}

\author{Bhagyashri Telsang$^1$, Seddik Djouadi$^1$, Charalambos D. Charalambous$^2$  
\thanks{$^{1}$Bhagyashri Telsang and Seddik Djouadi are  with the Department of Electrical Engineering and Computer Science, University of Tennessee, Knoxville, TN, 37996, USA 
{\tt\small \{btelsang, mdjouadi\}@utk.edu}}%
\thanks{$^{2}$Charalambos D. Charalambous is  with the Faculty of Electrical and Computer Engineering, University of Cyprus, Nicosia 1678, Cyprus
{\tt\small chadcha@ucy.ac.cy}}%
}

\begin{document}

\maketitle
\thispagestyle{empty}
\pagestyle{empty}

\begin{abstract}
Witsenhausen's 1968 counterexmaple  is a simple  two-stage  decentralized stochastic control problem that highlighted the difficulties of sequential  decision  problems with non-classical information structures. Despite extensive prior efforts, what is known currently, is the exact Person-by-Person (PbP) optimal nonlinear strategies, which satisfy two nonlinear integral equations,  announced in 2014, and obtained using Girsanov's change of measure transformations. In this paper, we provide numerical solutions to the two exact  nonlinear PbP optimal control strategies,  using the Gauss Hermite Quadrature to approximate the integrals and then solve a system of non-linear equations to compute the signaling levels. Further, we  analyse and compare our  numerical results to existing results  previously reported in the literature.

\end{abstract}

\section{INTRODUCTION}
\label{sec:introduction}

\begin{figure}
    \centering
    \includegraphics[width=\columnwidth]{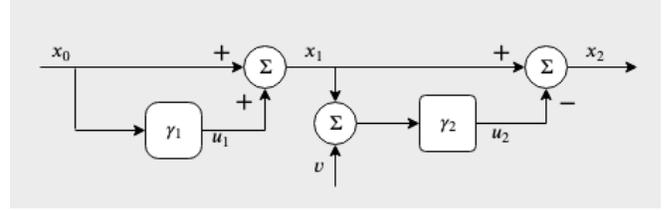}
    \caption{Witsenhausen's decentralized stochastic system}
    \label{fig:witsenhausenblkdia}
\end{figure}

The Witsenhausen's counterexample \cite{WitsenhausenOriginal} is a two-stage stochastic control problem, shown  in Fig. \ref{fig:witsenhausenblkdia},  described by the following (state and output) equations, admissible strategies and pay-off.

\noindent{\it State Equations:}
\begin{align}
    &x_1 = x_0 + u_1, \ x_0 \sim p_{x_0}(\cdot) \nonumber \\
    &x_2 = x_1 - u_2 \label{eq:stateeq}
\end{align}

\noindent{\it Output Equations:}
\begin{align}
y_0 &= x_0, \nonumber \\
y_1 &= x_1 +v, \ \hst v \sim \mathcal{N}(0,\sigma^2)  \label{eq:opeq}
\end{align}

\noindent{\it Admissible Borel Measurable Strategies:}
\begin{align}
u_1 = \gamma_1(y_0), \hst 
u_2 = \gamma_2(y_1)  \label{eq:admissiblestrategies}
\end{align}

\noi{\it Cost function:}
\begin{align}
J(u_1,u_2) = J(\gamma_1,\gamma_2) \triangleq {\mathbb E}\Big\{ k^2 u_1^2 + x_2^2 \Big\}, \ k\in\mathbb{R} \label{eq:costeq}
\end{align}
Here, $x_0 :\Omega \rar {\mathbb R}$ is a random variable (RV) with known probability density function $p_{x_0}(\cdot)$, and is independent of $v$. Without loss of generality, we consider $x_0 \sim \mathcal{N}(0,\sigma_x^2)$.

The main objective of the Witsenhausen counterexample is to determine a tuple of strategies $(\gamma_1^*, \gamma_2^*)$ that minimize $J(\gamma_1,\gamma_2)$, i.e.,
\begin{align}
J(\gamma_1^*, \gamma_2^*) \triangleq & \inf_{(\gamma_1(y_0), \gamma_2(y_1))}  J(\gamma_1,\gamma_2) .  \label{global}
\end{align}
The exact form of the nonlinear strategies $(\gamma_1^*, \gamma_2^*)$ is currently unkown; the difficulty is attributed to the fact that,    $y_0$ is known to the control strategy $\gamma_1$ but not to the control strategy $\gamma_2$, i.e., the information structure is nonclassical  \cite{WitsenhausenOriginal}.

\subsection{Prior Literature}
 Hans Witsenhausen in \cite{WitsenhausenOriginal},  analyzed the counterexample extensively; he showed that  optimal strategies exists,  and for certain parameters $(k, \sigma^2)$,  constructed a sub-optimal tuple of nonlinear strategies that  outperform the tuple of  optimal affine or linear strategies (these are recalled  in Section \ref{subsec:results}, see  (\ref{affine}) and  \eqref{eq:WitsenhausenNonlinearLaws}). We should emphasize that  Theorem 2 of \cite{WitsenhausenOriginal} does not claim  that nonlinear strategies  outperform all affine strategies for all values of parameters $(k, \sigma^2)$; rather,  it is only for certain parameters that the sub-optimal nonlinear strategies  \eqref{eq:WitsenhausenNonlinearLaws}  outperform the  optimal affine strategies (\ref{affine}). \\
 One of the main results of  Witsenhausen is: for a fixed $\gamma_1$ the  optimal strategy   $\gamma_2^*(y_1)$ is  \cite{WitsenhausenOriginal}:
\begin{equation}
    \gamma_2^*(y_1) = \mathbb{E}\{\bar{\gamma}_1(x_0) | y_1\}, \hst \bar{\gamma}_1(x_0)= x_0 + \gamma_1(x_0). \label{wit_s_2}
\end{equation}
However, the optimal strategy $\gamma_1^*(y_1)$ is currently unknown.  
 
The Witsenhausen's counterexample received much attention over the years by the  control and information theory communities.   \cite{BansalWhenisAffineOptimal} parameterized the tuple of strategies $(\gamma_1, \gamma_2)$ by partitioning the parameter space into two regions: one with an  affine strategy  and the other with a nonlinear strategy.

  \cite{FEanalysisRomvary} applied finite element methods to develop numerical schemes  to compute the optimal pay-off, when $\gamma_2^*$ is  given by (\ref{wit_s_2}). 
   \cite{neuralnetworksolution} developed  a numerical scheme to compute the pay-off by employing one-hidden-layer neural network, making use of $\gamma_2^*$  given by (\ref{wit_s_2}).  \cite{iterativecodingWit} applied approaches an  iterative source-channel coding method to quantize the strategies.  \cite{hierarchialLee} developed  numerical  methods based on  nonconvex optimization.  \cite{McEany2} applied  making use of $\gamma_2^*$  given by (\ref{wit_s_2}), and   transformed the problem to an optimization problem over the space of quantile functions, and provided a numerical scheme that   generates the  approximate pay-off. 
   
Charalambous and Ahmed \cite{Bambos_equivalence} in 2014 computed the exact nonlinear strategies $(\gamma_1^o,\gamma_2^o)$ using the notion of Person-by-Person (PbP) optimality. The approach in   \cite{Bambos_equivalence} is the formulation of an  equivalent optimization problem, under a reference probability measure, such that the observation $y_1$ is independent of the strategies    $(\gamma_1, \gamma_2)$. This approach is fully described in \cite{Bambos_equivalence,charalambous:MCSS2016,charalambous-ahmed:IEEEAC2017a} for decentralized problems described by stochastic differential equations. \\ 
The exact optimal PbP strategies are given by \cite{Bambos_equivalence}
\begin{align}
\gamma_1^o(x_0) &= -\frac{1}{2k^2\sigma^2} \mathbb{E} \Big\{ (y_1-\bar{\gamma}_1^o(x_0)) (\bar{\gamma}_1^o(x_0) - \gamma_2^o(y_1))^2 \Big\} \nonumber \\
&- \frac{1}{k^2}\mathbb{E}\Big\{  \bar{\gamma}_1^o(x_0) - \gamma_2^o(y_1) |x_0 \Big\} \label{law_1}
\end{align}
\begin{align}
\gamma_2^o(y_1)={\mathbb E}\{\bar{\gamma}_1^o(y_1)|x_0\}, \hso  \bar{\gamma}_1^o(y_1)= x_0   +   \gamma_1^o(x_0)  \label{law_2} . 
\end{align}
The above PbP optimal strategies are equivalently expressed in terms of $\bar{\gamma}_1^o=x_0+\gamma_1^o$ and $\gamma_2^o$  by 
 the two nonlinear integral equations:
\begin{align}
&    \bar{\gamma}_1^o(x_0) = x_0 - \frac{1}{k^2} \bigintssss_{-\infty}^{\infty} \Big\{\frac{1}{2\sigma^2} (\zeta - \bar{\gamma}_1^o(x_0)) (\bar{\gamma}_1(x_0) - \gamma_2^o(\zeta))^2 \nonumber \\ 
 &   + (\bar{\gamma}_1^o(x_0) - \gamma_2^o(\zeta))\Big\} \frac{1}{\sqrt{2\pi \sigma^2}} \exp({-\frac{(\zeta-\bar{\gamma}_1^o(x_0))^2}{2\sigma^2}}) d\zeta
    \label{eq:gamma1bar}
\end{align}
\begin{align}
    \gamma_2^o(y_1) &= \frac{\bigintssss_{-\infty}^{\infty} \bar{\gamma}_1^o(\xi) \exp{(-\frac{(y_1-\bar{\gamma}_1^o(\xi))^2}{2\sigma^2})}  p_{x_0}(d\xi)}{\int_{-\infty}^{\infty} \exp{(-\frac{(y_1-\bar{\gamma}_1^o(\xi))^2}{2\sigma^2})} p_{x_0}(d\xi)}
    \label{eq:gamma2}
\end{align}
It is important to mention that the PbP strategy $\gamma_2^o$, i.e.,  (\ref{law_2}) is derived by applying PbP optimality, i.e., calculus of variations, and that $\gamma_2^o$ has the form derived by Witsenhausen in \cite{WitsenhausenOriginal}, i.e., (\ref{wit_s_2}). That is, the functional forms satisfy $\gamma_2^o=\gamma_2^*$. However, we do not yet know, if Witsenhausen's global optimal strategy $\gamma_1^*$ is identlical to the PbP optimal strategy $\gamma_1^o$. 

\subsection{Contributions of the Paper}
In this paper we undertake the study of calculating the optimal PbP strategies, by approximating the integrals \eqref{eq:gamma1bar} and \eqref{eq:gamma2}, using  the Gauss Hermite Quadrature numerical integration method. The resulting coupled approximations are then solved by posing them as a system of nonlinear equations; this method is detailed in Section \ref{sec:NumericalImplementation}.

One of the main contributions is to evaluate the performance of the  exact nonlinear PbP strategies with respect to the properties of the global optimal strategies $(\gamma_1^*, \gamma_2^*)$ derived by Witsenhausen in  \cite{WitsenhausenOriginal}.

The findings are presented in Section \ref{subsec:results}, for different  parameter values $(k, \sigma^2)$.  The conclusions are found in Section \ref{sec::conclusion}.

\section{Numerical Integration of the Optimal Strategies}
\label{sec:NumericalImplementation}


Consider the optimal strategies in their integral form \eqref{eq:gamma1bar} and \eqref{eq:gamma2}. Recognizing that with the exponential function within the integral, the integral form can be reformulated to have a Gaussian exponential function, we employ the Gauss Hermite Quadrature (GHQ) method to implement the optimal strategies.

First, we briefly review the Gauss Hermite Quadrature method. The approximate numerical integration formula for a function $f(x)$ on the infinite range $(-\infty,\infty)$ with the weight function $e^{-x^2}$ is \cite{GHQGreenwood}:

\begin{equation}
    \int_{-\infty}^{\infty} f(x) e^{-x^2} dx \approx \Sigma_{i=1}^n f(x_{i,n}) \lambda_{i,n}
    \label{eq:GaussQuadratureRule}
\end{equation}

\noindent where the abscissas $\{x_{i,n}\}$ are the roots of the $n^{th}$ order Hermite polynomial

\begin{align*}
    H_n(x) = -\sqrt{2}^n h_n( \sqrt{2}x) = 0 
\end{align*}

with $h_n(x) = e^{\frac{x^2}{2}} \frac{d^n(e^{\frac{-x^2}{2}})}{dx^n}$ and the weights $\{\lambda_{i,n}\}$ are given by

\begin{align*}
    \lambda_{i,n} = \frac{\sqrt{\pi}2^{n+1}{n!}}{H_n^{'}(x_{i,n})^2}
\end{align*}

\noindent where $H_n^{'}(x) = 2nH_{n-1}(x)$. For $n\leq 10$, the zeros $x_{i,n}$ of the Hermite polynomial $H_n(x)$ and the weights $\lambda_{i,n}$ are calculated in \cite{GHQGreenwood}. For higher orders, the zeros and weights are calculated in \cite{MaxwellGHQ}. It is shown in \cite{GolubGHQ} that the Gauss quadrature rule \eqref{eq:GaussQuadratureRule} is exact for all continuous functions $f$ that are polynomials of degree $\leq 2n-1$. The implications of quadrature rule to approximate a discontinuous function will be discussed in Section \ref{subsec:toughparameters}.

It is in general a difficult problem to compute zeros and weights for any Hermite polynomial and any weight function. Therefore, since the zeros and weights for the aforementioned $H_n(x)$ are calculated in the literature, we transform the optimal strategies (\ref{eq:gamma1bar}) and (\ref{eq:gamma2}) to have the standard Gaussian function $e^{-x^2}$ as the weight function.

Consider the first law (\ref{eq:gamma1bar}) and the change of variables as $z = {\frac{\zeta-\bar{\gamma}_1^o(x_0)}{\sqrt{2\sigma^2}}}$ and $du = {\frac{d\zeta}{\sqrt{2\sigma^2}}}$. Then,

\begin{equation*}
    \begin{split}
        \bar{\gamma}_1^o(x_0) = x_0 - \frac{1}{\sqrt{\pi}k^2} \bigintssss_{-\infty}^{\infty} \Big\{\frac{z}{\sqrt{2\sigma^2}} \\ (\bar{\gamma}_1^o(x_0) - 
        \gamma_2^o(\sqrt{2\sigma^2}z + \bar{\gamma}_1^o(x_0)))^2 \\ +  (\bar{\gamma}_1^o(x_0) - \gamma_2^o(\sqrt{2\sigma^2}z + \bar{\gamma}_1^o(x_0)))\Big\} e^{-z^2} dz
    \end{split}
\end{equation*}
Using Gauss Hermite Quadrature approximation \eqref{eq:GaussQuadratureRule},
\begin{equation}
    \begin{split}
        \bar{\gamma}_1^o(x_0) \approx x_0 - \frac{1}{\sqrt{\pi}k^2} \sum_{i=1}^n \Big\{\frac{z_i}{\sqrt{2\sigma^2}} \\
        (\bar{\gamma}_1^o(x_0) - \gamma_2^o(\sqrt{2\sigma^2}z_i + \bar{\gamma}_1^o(x_0)))^2 \\
        + (\bar{\gamma}_1^o(x_0) - \gamma_2^o(\sqrt{2\sigma^2}z_i + \bar{\gamma}_1^o(x_0)))\Big\} \lambda_i
    \end{split}
    \label{eq:gamma1bar_GHQ}
\end{equation}
Similarly approximating the second law (\ref{eq:gamma2}) with the change of variable $z = \frac{\xi}{\sqrt{2\sigma_x^2}}$, we get:

\begin{align}
    \gamma_2^o(y_1) = \frac{\bigintssss_{-\infty}^{\infty} \bar{\gamma}_1^o(\xi) \exp{(-\frac{(y_1-\bar{\gamma}_1^o(\xi))^2}{2\sigma^2})}  \exp{(-\frac{\xi^2}{2\sigma_x^2})} d\xi}{\int_{-\infty}^{\infty} \exp{(-\frac{(y_1-\bar{\gamma}_1^o(\xi))^2}{2\sigma^2})} \exp{(-\frac{\xi^2}{2\sigma_x^2})} d\xi} 
    \label{eq:gamma2GHQ}
\end{align}

Consider \eqref{eq:gamma1bar_GHQ}, since $z_i$ and $\lambda_i$ are the (known) nodes and weights, for certain $x_0 \in \mathbb{R}$, the unknowns are $\bar{\gamma}_1^o(x_0)$ and $\gamma_2^o(\sqrt{2\sigma^2}z_i + \bar{\gamma}_1^o(x_0)))$ (whose argument is in turn a function of $\bar{\gamma}_1^o(x_0)$). In order to solve this equation, we employ the expression for $\gamma_2^o(y_1)$ from \eqref{eq:gamma2GHQ} by having $y_1 = \sqrt{2\sigma^2}z_i + \bar{\gamma}_1^o(x_0)$. Substituting $\gamma_2^o(y_1=\sqrt{2\sigma^2}z_i + \bar{\gamma}_1^o(x_0)))$ from \eqref{eq:gamma2GHQ} in (\ref{eq:gamma1bar_GHQ}) to get:
\begin{align}
    \begin{split}
       & \bar{\gamma}_1^o(x_0) \approx x_0 - \frac{1}{\sqrt{\pi}k^2} \sum_{i=1}^n \lambda_i \Bigg\{\frac{z_i}{\sqrt{2\sigma^2}} \\ &\Bigg(\bar{\gamma}_1^o(x_0) - 
        \bigg( \sum_{j=1}^{n} \big( \bar{\gamma}_1^o(\sqrt{2\sigma_x^2} z_j)  \\
      &\exp{(-\frac{(\sqrt{2\sigma^2}z_i + \bar{\gamma}_1^o(x_0)-\bar{\gamma}_1^o(\sqrt{2\sigma_x^2} z_j))^2}{2\sigma^2})} \lambda_j \big) \bigg) \bigg/  \\
      &\bigg( \sum_{j=1}^{n} \Big( \exp{(-\frac{(\sqrt{2\sigma^2}z_i + \bar{\gamma}_1^o(x_0)-\bar{\gamma}_1^o(\sqrt{2\sigma_x^2} z_j))^2}{2\sigma^2})} \lambda_j \Big) \bigg)\Bigg)^2 \\
       & + \Bigg(\bar{\gamma}_1^o(x_0) - \bigg( \sum_{j=1}^{n} \big( \bar{\gamma}_1^o(\sqrt{2\sigma_x^2} z_j) \\
      &\exp{(-\frac{(\sqrt{2\sigma^2}z_i + \bar{\gamma}_1^o(x_0)-\bar{\gamma}_1^o(\sqrt{2\sigma_x^2} z_j))^2}{2\sigma^2})} \lambda_j \big) \bigg) \bigg/ \\
      &\bigg( \sum_{j=1}^{n} \Big( \exp{(-\frac{(\sqrt{2\sigma^2}z_i + \bar{\gamma}_1^o(x_0)-\bar{\gamma}_1^o(\sqrt{2\sigma_x^2} z_j))^2}{2\sigma^2})} \lambda_j \Big) \bigg)\Bigg)\Bigg\} 
    \end{split}
    \label{eq:bignonlineareq}
\end{align}

While $x_0 \in \mathbb{R}$ and $\sqrt{2\sigma_x^2} z_i$ are known, $\bar{\gamma}_1^o(x_0)$ and $\bar{\gamma}_1^o(\sqrt{2\sigma_x^2} z_i)$ are unknown. Let $s_i = \bar{\gamma}_1^o(\sqrt{2\sigma_x^2} z_i), \ \forall i$. For each $x_0$, \eqref{eq:bignonlineareq} hence contains $(n+1)$ number of unknowns, i.e., $n$ $s_i$'s and one $\bar{\gamma}_1^o(x_0)$:
\begin{align*}
    & \bar{\gamma}_1^o(x_0) \approx x_0 - \frac{1}{\sqrt{\pi}k^2} \sum_{i=1}^n \lambda_i \Bigg\{\frac{z_i}{\sqrt{2\sigma^2}} \\ &\Bigg(\bar{\gamma}_1^o(x_0) - 
        \bigg( \sum_{j=1}^{n} \big( s_j 
      \exp{(-\frac{(\sqrt{2\sigma^2}z_i + \bar{\gamma}_1^o(x_0)-s_j)^2}{2\sigma^2})} \lambda_j \big) \bigg) \bigg/  \\
      &\bigg( \sum_{j=1}^{n} \Big( \exp{(-\frac{(\sqrt{2\sigma^2}z_i + \bar{\gamma}_1^o(x_0)-s_j)^2}{2\sigma^2})} \lambda_j \Big) \bigg)\Bigg)^2 +\\
       & \Bigg(\bar{\gamma}_1^o(x_0) - \bigg( \sum_{j=1}^{n} \big( s_j
      \exp{(-\frac{(\sqrt{2\sigma^2}z_i + \bar{\gamma}_1^o(x_0)-s_j)^2}{2\sigma^2})} \lambda_j \big) \bigg) \bigg/ \\
      &\bigg( \sum_{j=1}^{n} \Big( \exp{(-\frac{(\sqrt{2\sigma^2}z_i + \bar{\gamma}_1^o(x_0)-s_j)^2}{2\sigma^2})} \lambda_j \Big) \bigg)\Bigg)\Bigg\} 
\end{align*}

Substituting $x_0 = x_{0l} = \sqrt{2\sigma_x^2} z_l$ for each $l \in \{1,2,\hdots,n\}$, we obtain $n$ nonlinear equations with $n$ $s_l$'s that are unknown, given in \eqref{eq:sysnonlineareq}. Each $s_l$, which is the value of $\bar{\gamma}_1^o(x_0)$ at nodes selected according to Gauss-Hermite Quadrature, is the signaling level of the control action. Rearranging \eqref{eq:sysnonlineareq} to move all terms on one side, we denote the resulting system of nonlinear equations as $f_{sysnonlin}:\mathbb{R}^n \to \mathbb{R}^n$.

\begin{align}
    & \forall l = {1,2,\hdots,n} \nonumber \\
    & t_l \approx \sqrt{2\sigma_x^2} z_l - \frac{1}{\sqrt{\pi}k^2} \sum_{i=1}^n \lambda_i \Bigg\{\frac{z_i}{\sqrt{2\sigma^2}} \nonumber \\ &\Bigg(t_l - 
        \bigg( \sum_{j=1}^{n} \big( t_j 
      \exp{(-\frac{(\sqrt{2\sigma^2}z_i + t_l-t_j)^2}{2\sigma^2})} \lambda_j \big) \bigg) \bigg/ \nonumber \\
      &\bigg( \sum_{j=1}^{n} \Big( \exp{(-\frac{(\sqrt{2\sigma^2}z_i + t_l-t_j)^2}{2\sigma^2})} \lambda_j \Big) \bigg)\Bigg)^2 + \nonumber \\
       & \Bigg(t_l - \bigg( \sum_{j=1}^{n} \big( t_j
      \exp{(-\frac{(\sqrt{2\sigma^2}z_i + t_l-t_j)^2}{2\sigma^2})} \lambda_j \big) \bigg) \bigg/ \nonumber \\
      &\bigg( \sum_{j=1}^{n} \Big( \exp{(-\frac{(\sqrt{2\sigma^2}z_i + t_l-t_j)^2}{2\sigma^2})} \lambda_j \Big) \bigg)\Bigg)\Bigg\}
      \label{eq:sysnonlineareq}
\end{align}


The solution of the system of $n$ nonlinear equations \eqref{eq:sysnonlineareq} results in $n$ explicit points, i.e., $n$ signaling levels $s_l^*, \ \forall l=1,2,\hdots ,n,$ such that $||f_{sysnonlin}(s_1^*,s_2^*, \hdots , s_n^*)|| $ is close to zero. Using these $n$ signaling levels, we obtain the value of $\bar{\gamma}_1^o(x_0), \ \forall x_0,$ by substituting $(s_1^*,s_2^*, \hdots , s_n^*)$ in \eqref{eq:bignonlineareq} which results in one unknown $\bar{\gamma}_1^o(x_0)$ and solving the resulting nonlinear equation for $\bar{\gamma}_1^o(x_0)$ for each $x_0$. This is similar to the collocation method used to solve integral equations, \cite{PiecewiseCollocationAtkinson}. Here,  $x_0 = x_{0l} = \sqrt{2\sigma_x^2} z_l$ for each $l \in \{1,2,\hdots,n\}$ are the collocation points and signaling levels are the values of $\bar{\gamma}_1^o(x_0)$ at the collocation points. 

To obtain the strategy of the second controller, we substitute the signaling levels $(s_1^*,s_2^*, \hdots , s_n^*)$ in \eqref{eq:gamma2GHQ}. This directly gives the expression for $\gamma_2^o(y_1)$ which is evaluated at $y_1$. It is worth noting here that although $y_1 \in \mathbb{R}$, but because $y_1 = \bar{\gamma}_1^o(x_0) + v$ from \eqref{eq:opeq}, the values taken by $y_1$ are dictated by the strategy of the first controller $\bar{\gamma}_1^o(x_0)$. Once both the strategies $\bar{\gamma}_1^o$, $\gamma_2^o$ are obtained, we calculate the total cost $J$ from \eqref{eq:costeq}. 

We now briefly summarize the methodology to numerically integrate the derived optimal strategies \eqref{eq:gamma1bar} and \eqref{eq:gamma2}. 

\begin{framed}
\noindent Input parameters: $k, \sigma, \sigma_x, n$ \\
Input signals: $x_0, v$
\begin{itemize}
    \item[-] Solve $f_{sysnonlin}$ to obtain the signaling levels $(s_1^*,s_2^*, \hdots , s_n^*)$
    \item[-] For each $x_0$, compute $\bar{\gamma}_1^o(x_0)$
    \item[-] For all $y_1  = \bar{\gamma}_1^o(x_0) + v$, compute $\gamma_2^o(y_1)$
\end{itemize}
\end{framed}



\section{Results}
\label{subsec:results}

We employ the software MATLAB to implement the solution strategies \eqref{eq:gamma1bar} and \eqref{eq:gamma2}. The command \textit{fsolve} is used to solve the system of nonlinear equations $f_{sysnonlin}$ and \textit{lsqnonlin} to solve for $\bar{\gamma}_1^o(x_0)$. 

The set of parameters in the Witsenhausen counterexample \eqref{eq:stateeq}-\eqref{eq:costeq} are $(k,\sigma,\sigma_x)$. For certain sets of values of these parameters, the optimal law is affine while for the rest of the region of parameters, the optimal law is non-linear. In Lemma 1 in \cite{WitsenhausenOriginal}, Witsenhausen derived the optimal affine laws as:

\begin{align}
    \bar{\gamma}_1^{aff}(x_0) = \nu x_0 \nonumber \\
    \gamma_2(y_1)^{aff} = \mu y_1  \label{affine}
\end{align}
where $\bar{\gamma}_1(x_0) = x_0 + \gamma_1(x_0)$,
\begin{equation*}
    \mu = \frac{\sigma_x^2 \nu^2}{1+\sigma_x^2 \nu^2}
\end{equation*}
and $t=\sigma_x\nu$ is a real root of the equation
\begin{equation*}
    (t-\sigma_x)(1+t^2)^2 + \frac{1}{k^2}t = 0
\end{equation*}

We denote the cost obtained from the optimal affine laws as $J^{aff} = J(\bar{\gamma}_1^{aff},\gamma_2^{aff})$. In Theorem 2 of \cite{WitsenhausenOriginal} he considers the sample non-linear laws:

\begin{align}
    \bar{\gamma}_1^{wit}(x_0) &= \sigma_x \text{sgn}(x_0) \nonumber \\
    \gamma_2^{wit}(y_1) &= \sigma_x \tanh{(\sigma_xy_1)}
    \label{eq:WitsenhausenNonlinearLaws}
\end{align}

\noindent and shows that $J^{wit} < J^{aff}$ as $k \to 0$, where $J^{wit}$ is the cost resulting from the nonlinear laws \eqref{eq:WitsenhausenNonlinearLaws}. 

We denote the cost we obtain from the derived optimal laws \eqref{eq:gamma1bar} and \eqref{eq:gamma2} and implemented using the Gauss Hermite Quadrature numerical integration method detailed in Section \ref{sec:NumericalImplementation} as $J^o$. We consider different parameter values of $(k, \sigma, \sigma_x)$ and compare the cost we obtain $J^o$ with $J^{aff}$, $J^{wit}$ and some other costs previously reported in the literature. For additional insight into the results, the total cost is broken into two stages: Stage 1 and Stage 2 costs are the first and the second term, respectively, in the total cost:

\begin{equation}
    J(\bar{\gamma}_1^o,\gamma_2^o) = \mathbb{E}\big\{k^2 (\bar{\gamma}_1^o(x_0) - x_0)^2 + (\bar{\gamma}_1^o(x_0) - \gamma_2^o(y_1))^2 \big\} \label{eq:TotalCost}
\end{equation}

We have employed $600,000$ samples for $x_0$ and $v$ generated according to $\mathcal{N}(0,\sigma_x)$ and $\mathcal{N}(0,\sigma)$ respectively. The order of the Hermite polynomial in GHQ method is $n=7$. As stated in Lemma 1 of \cite{WitsenhausenOriginal}, the optimal cost is less than min$(1,k^2\sigma_x^2)$. Accordingly, we verify if the cost $J^o$ is less than min$(1,k^2\sigma_x^2)$.

\subsubsection{\textbf{Parameters $k=0.001, \sigma_x=1000, \sigma = 1$}}

\begin{figure}
    \centering
    \includegraphics[width=\columnwidth]{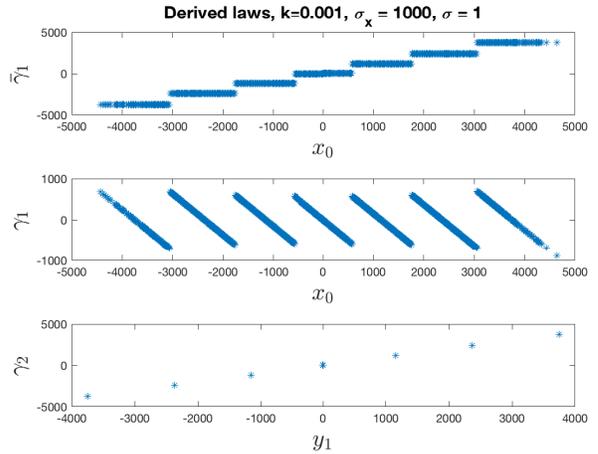}
    \caption{Optimal PbP strategies based on  \eqref{eq:gamma1bar} and \eqref{eq:gamma2}.}
    \label{fig:k0p001sigmax1000}
\end{figure}

The total costs obtained are reported in Table \ref{tab:k0p001sigmax1000}. Note that $J^o < \text{min}(1,k^2\sigma_x^2) = 1$ and so are $J^{wit}$ and $J^{aff}$. The optimal PbP strategies $\bar{\gamma}_1^o, \gamma_1^o$ and $\gamma_2^o$ obtained are shown in Fig. \ref{fig:k0p001sigmax1000}. As pointed in \cite{WitsenhausenOriginal}, we observe that PbP $\bar{\gamma}_1^o$ is indeed symmetric around  the origin. Moreover, we obtain four signaling levels, compared to one resulting from $\bar{\gamma}_1^{wit}$. We also observe that the derived strategies  result in a strategy such that $\bar{\gamma}_1^o(x_0) \approx \gamma_2^o(y_1)$ leading to near zero Stage 2 cost. It is worth pointing out that since $\gamma_2^o$ admits $y_1 = \bar{\gamma}_1^o(x_0) + v$ as the input, the behaviour of $\gamma_2^o$ over the entire real line $\mathbb{R}$ is not apparent. Moreover, despite the high value of $\sigma_x=1000$, the presented methodology is not numerically unstable. 

\begin{table}[]
\centering
\begin{tabular}{@{}cccc@{}}
\toprule
          & Stage 1  & Stage 2                 & Total Cost \\ \midrule
$J^{aff}$ & $0.9984$ & $9.9843 \times 10^{-7}$ & $0.9984$   \\
$J^{wit}$ & $0.4041$ & $0$                     & $0.4041$   \\
$J^o$  & $0.1137$ & $1.1368 \times 10^{-7}$ & $0.1137$   \\ \bottomrule
\end{tabular}
\caption{Total cost, $k=0.001$, $\sigma_x=1000$}
\label{tab:k0p001sigmax1000}
\end{table}

\subsubsection{\textbf{Parameters $k=1, \sigma_x=1, \sigma = 1$}}
\label{subsec:parameters111}
\begin{figure}
    \centering
    \includegraphics[width=\columnwidth]{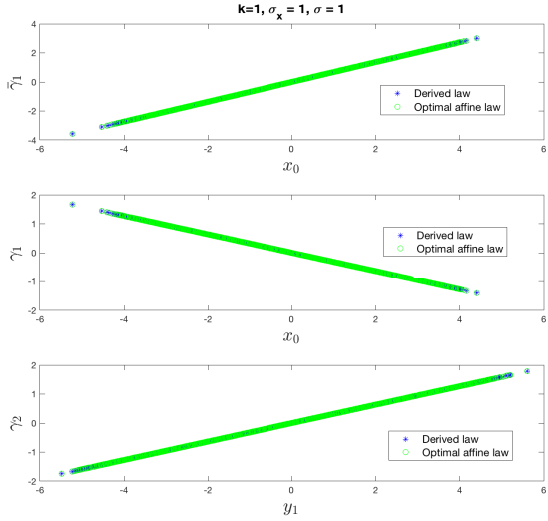}
    \caption{{Comparison of the optimal PbP strategies  and the  optimal affine strategies}}
    \label{fig:k1sigmax1}
\end{figure}

As pointed in \cite{WuVerduTransport}, this set of parameter values ($k \nless 0.56$ and $\sigma_x$ is not large) is in the region where affine laws are optimal. The optimal control laws \eqref{eq:gamma1bar} and \eqref{eq:gamma2} are compared with optimal affine laws in Fig. \ref{fig:k1sigmax1}. It is seen that the resulting laws are almost the same as the optimal affine laws. We further compare the cost with $J^{aff}$ and $J^{wit}$ in Table \ref{tab:k1sigmax1}. The negligible difference in $J^{aff}$ and $J^o$ is attributed to numerical inaccuracy in the implementation of \eqref{eq:gamma1bar} and \eqref{eq:gamma2} through approximate numerical integration method.

\begin{table}
\centering
\begin{tabular}{@{}cccc@{}}
\toprule
          & Stage 1  & Stage 2  & Total Cost \\ \midrule
$J^{aff}$ & $0.1011$ & $0.3174$ & $0.418500414352474$   \\
$J^{wit}$ & $0.4043$ & $0.4480$ & $0.852287449358227$   \\
$J^o$  & $0.1011$ & $0.3174$ & $0.418500469701766$   \\ \bottomrule
\end{tabular}
\caption{Total cost, $k=1, \sigma_x=1$}
\label{tab:k1sigmax1}
\end{table}

\subsubsection{\textbf{Comparison with \cite{BasarVariations}}}

\begin{figure}
    \centering
    \includegraphics[width=\columnwidth]{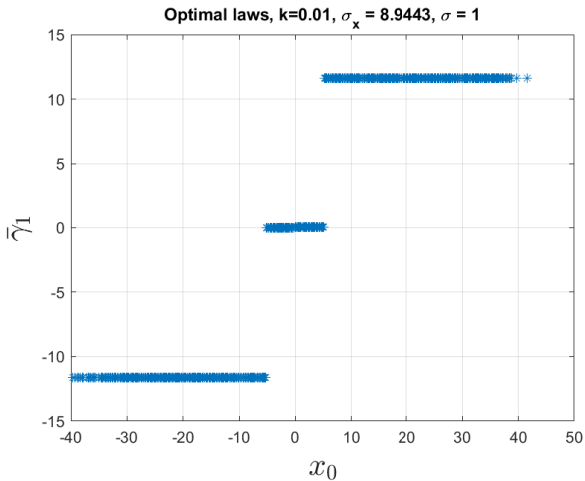}
    \caption{Optimal PbP strategies for the parameters in \cite{BasarVariations}}
    \label{fig:basarvariations}
\end{figure}

A class of nonlinear policies initially introduced in \cite{WitsenhausenOriginal} and further analyzed and improved upon in \cite{BansalWhenisAffineOptimal} is given by:

\begin{align}
    \gamma_1^{bb}(x_0) &= \epsilon^{bb} \text{sgn}(x_0) + \lambda^{bb} x_0 \nonumber \\
    \gamma_2^{bb}(y_1) &= \mathbb{E}[\epsilon^{bb} \text{sgn}(x_0) + \lambda^{bb}x_0 | y_1]
    \label{eq:bansalbasarnonlinearlaws}
\end{align}

\noindent where $\epsilon^{bb}$ and $\lambda^{bb}$ are parameters to be optimized over. For $k=0.01, \sigma_x = \sqrt{80}$ and $\sigma=1$, \cite{BasarVariations} picks $\epsilon^{bb}=5$ and $\lambda^{bb}=0.01006$ in the law \eqref{eq:bansalbasarnonlinearlaws} and reports the cost to be $J^{bb} = 0.3309$. Furthermore, the authors in \cite{McEany1BadPaper} mention that they obtain the same cost of $0.3309$ with the algorithm developed therein. The optimal law $\bar{\gamma_1}(x_0)$ that we obtain from \eqref{eq:gamma1bar} is shown in Fig. \ref{fig:basarvariations}. The corresponding total cost is compared with $J^{bb}$, $J^{wit}$ and the optimal affine cost $J^{aff}$ in Table \ref{tab:BasarVariations}. The stage 2 cost from both $J^o$ and $J^{aff}$ is of the order $10^{-7}$ and from $J^{wit}$ it is $0$. 

\begin{table}
\centering
\begin{tabular}{@{}cc@{}}
\toprule
          & $k=0.01,\sigma=1$    \\
          & $\sigma_x=\sqrt{80}$ \\ \midrule
$J^{aff}$ & $0.007986277332674$             \\
$J^{wit}$ & $0.003232551870223$             \\
$J^{bb}$  & $0.3309$             \\
$J^o$  & $0.001566775786064$             \\ \bottomrule
\end{tabular}
\caption{Total cost obtained from different solutions}
\label{tab:BasarVariations}
\end{table}


\subsubsection{\textbf{Parameters $k=0.2, \sigma_x=5, \sigma=1$}}
\label{subsec:toughparameters}
The last set of parameters we consider has been the most studied case and has enabled more insights into the solution of the Witsenhausen counterexample. \cite{neuralnetworksolution} provides a numerical solution by employing one-hidden-layer neural network as an approximating network. The cost obtained therein is denoted $J^{nn}$. Lee, Lau and Ho present a hierarchial search approach in \cite{hierarchialLee}. Therein, they impose $\bar{\gamma}_1$ to be a non-decreasing, step function that is symmetric about the origin (a property derived by Witsenhausen). For a number of steps, they find the signaling levels (value of $\bar{\gamma}_1$ at the step) and the breakpoints ($x_0$ where the step change occurs). They also find that the cost objective is lower for slightly sloped steps than perfectly leveled steps. Through comparison of their costs for different number of steps, they find that $7-$step solution yields the lowest cost. The cost obtained in \cite{hierarchialLee} is denoted $J^{llh}$ here and the signaling levels therein are $s^{*} = \{0, \pm 6.5, \pm 13.2, \pm 19.9\}$.

In our work, the solution of \eqref{eq:sysnonlineareq} yields the signling levels $s^{**} = \{0, \pm 6.15, \pm 12.8, \pm 19.8\}$ and $||f_{sysnonlin}(s^{**})|| = 10^{-15}$ while $||f_{sysnonlin}(s^{*})|| = 0.7$. Following up on the notes from Section \ref{sec:NumericalImplementation}, the Gauss quadrature rule is not exact for the set of parameters $k=0.2, \sigma_x=5, \sigma=1$ because this parameter set lies in the region where the optimal laws are non-linear. Moreover, the optimal non-linear laws are not continuous; they are only piecewise continuous. As a result, the inaccuracy in the approximation using Gauss quadrature rule is apparent. The cost we obtain for signaling levels $s^*$ and $s^{**}$ are $J^o_* = 0.16$ and $J^o_{**} = 0.1712$ respectively. 

\begin{figure}
    \centering
    \includegraphics[width=\columnwidth]{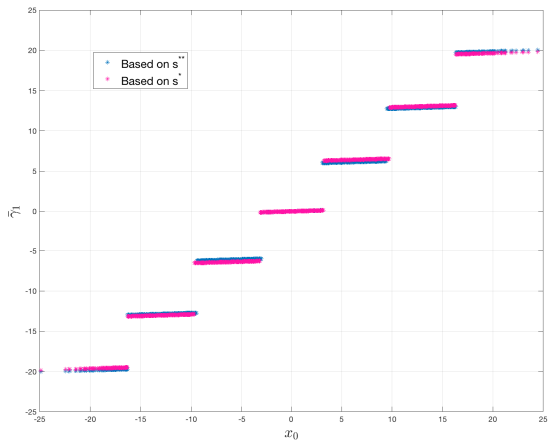}
    \caption{Pbp strategies for signaling levels $s^*$ and $s^{**}$}
    \label{fig:gamma1barsolcompare}
\end{figure}

\begin{figure}
    \centering
    \includegraphics[width=\columnwidth]{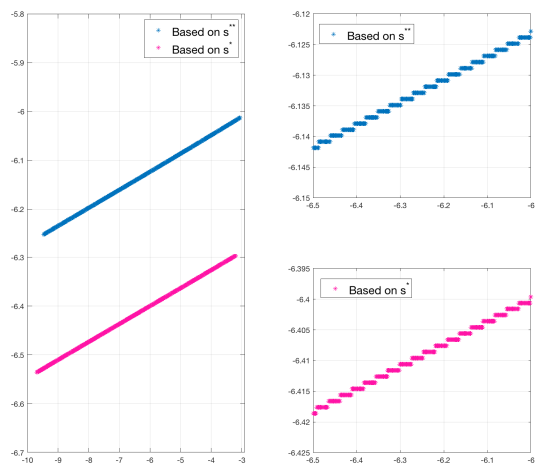}
    \caption{Signaling levels (magnified) are slightly sloped}
    \label{fig:slightlysloped}
\end{figure}

The optimal PbP strategy, $\bar{\gamma}_1^o(x_0)$, we have obtained for the signaling levels $s^*$ and $s^{**}$, are shown in Fig \ref{fig:gamma1barsolcompare}. Although we do not externally impose symmetry, it can be observed that $\bar{\gamma}_1^o$ is symmetric around the origin and is non-decreasing. We zoom in on one of the $7$ steps and observe in the left column of Fig \ref{fig:slightlysloped} that the steps are slightly sloped. Further zooming in, we see in the right column of Fig \ref{fig:slightlysloped} that each signaling level is further comprised of a number of closely spaced steps. Similar to this result, the authors in \cite{hierarchialLee} added segments in each of the $7$ steps to obtain the cost $J^{llh}=0.167313205338$. We compare both the costs with previously reported costs in the literature in Table \ref{tab:k0p2sigmax5}. Further in agreement with the findings in \cite{hierarchialLee}, we obtain the lowest cost for $7$ steps, $J^o_{**} = 0.1712$. 

\begin{table*}[t]
\centering
\begin{tabular}{@{}cccc@{}}
\toprule
          & Stage 1             & Stage 2                            & Total Cost          \\ \midrule
$J^{aff}$ & $0.0017428616051158$ & $0.956950417234115$                & $0.958693278839234$ \\
$J^{wit}$ & $0.403507741927546$ & $2.134488364684996 \times 10^{-6}$ & $0.403509876415911$ \\
$J^{nn}$\cite{neuralnetworksolution}  & -                   & -                                  & $0.1735$            \\
$J^{llh}$ \cite{hierarchialLee}  & $0.131884081844$        & $0.035429123524$                       & $0.167313205368$        \\
$J^o_*$  & $0.128541364988695$ & $0.038385613344897$                & $0.166926978333592$ \\
$J^{o}_{**}$  & $0.120110042087359$        & $0.051158481289032$                       & $0.171268523376388$        \\ \bottomrule
\end{tabular}
\caption{Reported and obtained costs, $k=0.2, \sigma_x=5, \sigma=1$}
\label{tab:k0p2sigmax5}
\end{table*}

With the parameter set $k=0.2, \sigma_x=5, \sigma=1$, the number of steps we obtain is the same as the value of the Gauss quadrature rule parameter $n$. However, this is not necessarily the case for all parameter sets; for example see Section \ref{subsec:parameters111}. The parameter set  $k=1, \sigma_x=5, \sigma=1$ is known to lie in a region where the optimal law is affine, and even though we employ $n=7$ order for GHQ, the resulting control laws are affine. Likewise, as seen in Fig \ref{fig:basarvariations}, the parameter set lies in the region where the optimal law is non-linear and we obtain a three-step control strategy  for $\bar{\gamma}_1^o$ for the GHQ order $n=7$.

\section{Conclusion}
\label{sec::conclusion}
Computed are the exact optimal PbP  strategies of the Witsenhausen counterexample derived in \cite{Bambos_equivalence}, that  satisfy the tuple of  nonlinear integral equations  \eqref{eq:gamma1bar} and \eqref{eq:gamma2}, using  the Gauss hermite quadrature scheme,  to transform the integral equations to a system of nonlinear equations. 

Comparison to  various costs  obtained in the literature show that strategies \eqref{eq:gamma1bar} and \eqref{eq:gamma2} outperform  previously reported results  for most parameters values. Moreover, PbP strategies  \eqref{eq:gamma1bar} and \eqref{eq:gamma2} reduce to optimal affine laws, for certain paremeters.

The computed optimal PbP  strategies are approximations of the exact  PbP optimal strategies, because our numerical scheme,  based on the Gauss Hermite Quadrature numerical integration,  is not exact,  when the underlying functions are not continuous. 

Since  the tuple of PbP optimal strategies \eqref{eq:gamma1bar} and \eqref{eq:gamma2} predict the properties of global optimal strategies $ (\gamma_1^*,\gamma_2^*)$ of the Witsenhausen problem defined by (\ref{global}), it is natural to investigate, in future work,  whether $(\gamma_1^o,\gamma_2^o)=(\gamma_1^*,\gamma_2^*)$.



\end{document}